\documentclass[epj]{svjour} 

\usepackage{bm}
\usepackage{amsmath}
\usepackage{wasysym}
\usepackage{amssymb}
\usepackage{graphics}
\usepackage{graphicx}
\usepackage{subfigure}
\usepackage{bbm}
\usepackage[english]{babel}

\begin{document}

\title{Atom interferometry gravity-gradiometer for the determination of the Newtonian gravitational constant
\textit{G}}
\titlerunning{Atom interferometry gravity-gradiometer for the determination of the
Newtonian...}

\author{ A. Bertoldi \and
G. Lamporesi \and L. Cacciapuoti\thanks{European Space Agency, ESTEC, 2200 AG
Nordwijk, NL} \and M. de Angelis\thanks{On leave from: Istituto Cibernetica
CNR,80078 Pozzuoli, I} \and M. Fattori\thanks{Present address: Physikalisches
Institut, Universit\"at Stuttgart, 70550 Stuttgart, D} \and T.
Petelski\thanks{Present address: European Patent Office,80469 M\"unchen,D} \and
A. Peters\thanks{Institut f\"ur Physik, Humboldt--Universit\"at zu Berlin,
10117 Berlin, D} \and M. Prevedelli\thanks{Dipartimento di Chimica Fisica e
Inorganica, Universit\`a di Bologna, 40136 Bologna, I} \and
J.~Stuhler${^{\text{c}}}$ \and G. M. Tino
\thanks{E-mail: guglielmo.tino@fi.infn.it} }

\institute{ Dipartimento di Fisica and LENS, Universit\`a di Firenze; INFN,
Sezione di Firenze \\ Via Sansone 1, 50019 Sesto Fiorentino (Firenze), Italy }

\date{Received: date / Revised version: date}

\abstract{We developed a gravity-gradiometer based on atom
  interferometry for the determination of the Newtonian
  gravitational constant \textit{G}. The apparatus, combining a Rb
  fountain,  Raman interferometry  and a juggling scheme
  for fast launch of two atomic clouds, was specifically designed
  to reduce possible systematic effects. We present
  instrument performances and show that the sensor is able to
  detect the gravitational field induced by source masses. A discussion of
  projected accuracy for \textit{G} measurement
  using this new scheme shows that the results of the
  experiment will be significant to discriminate between previous inconsistent
  values.}

\PACS {
  {03.75.Dg}{Atom and neutron interferometry} \and
  {04.80.--y}{Experimental studies of gravity} \and
  {06.20.Jr}{Determination of fundamental constants}}

\maketitle

\hyphenation{A-to-mi-ca} \hyphenation{grouped} \hyphenation{sta-bi-lized}
\hyphenation{puls-es} \hyphenation{Dop-pler} \hyphenation{switched}
\hyphenation{points}

\section{Introduction}

Recent advances in atom interferometry
led to the development of new methods for fundamental physics experiments and
for applications \cite{berman:ai}. In particular, atom interferometers are new
tools for experimental gravitation  as, for example, for precision measurements
of gravity acceleration \cite{chu:gravimeter}, gravity gradients
\cite{kasevich:gradiometer1},
  equivalence principle tests \cite{haensch:EP},
$1/r^{2}$ law test \cite{tino01bis,dimopulos03,ferrari:Sr},
 and for possible applications in geophysics
\cite{chu:gravimeter,kasevich:gradiometer2}. Ongoing studies show that future
experiments in space \cite{tino01bis} will allow to take full advantage of the
potential sensitivity of atom interferometers using, for example, atom
gyroscopes \cite{kasevich:gyro,landragin:int} to test general relativity
predictions \cite{HYPER}. The possibility of using atom interferometry for
gravitational waves detection was also investigated
\cite{tino:GW,chiao:GW,roura06}.

In this paper, we describe an atom interferometer developed for a precise
determination of the Newtonian gravitational constant \textit{G}. The basic
idea of the experiment and the planned precision were  presented in
\cite{fattori:magia}. Here we discuss the interferometer performances and show
that the sensor is able to  detect the gravitational field induced by source
masses. The projected accuracy for \textit{G} measurement using this new scheme
shows that the results of the  experiment will be significant to discriminate
between existing inconsistent  values.

In fact, because of the importance of this fundamental physical constant, more
than 300 experiments were performed to determine its value but the results are
not in agreement.  As a result, the present recommended CODATA value
(\textit{G}~=~6.6742(10)$\times$$10^{-11}$
$\text{m}^{3}\text{kg}^{-1}\text{s}^{-2}$)  is affected by an uncertainty of
150 ppm, which is  much larger than for any other fundamental physical constant
\cite{CODATA}.
With a few exceptions, most experiments were performed using conceptually
similar schemes based on suspended macroscopic masses as probes and torsion
balances or pendula as detectors.

\begin{figure}[t]
\vspace{-0mm}
\begin{center}
\hspace{-0mm}
\includegraphics[width=0.45\textwidth,angle=0]{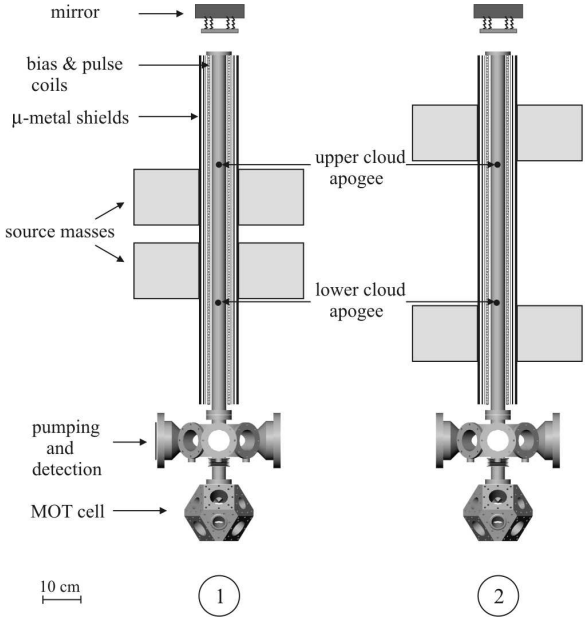}
\vspace{0mm} \caption{\label{fig:setup} Scheme of the experimental apparatus
showing the atomic fountain set-up and the two source masses configurations.
The laser beams and optical system are not shown.}
\end{center}
\end{figure}

%

In our experiment, freely falling atoms act as probes of the gravitational
field and an atom interferometry scheme is used to measure  the effect of
nearby well-characterized source masses (Figure \ref{fig:setup}).
${^{87}\text{Rb}}$ atoms, trapped and cooled in a magneto--optical trap (MOT),
 are launched upwards in a vertical vacuum tube
with a moving optical molasses scheme, producing an atomic fountain.
 Near the apogee of the atomic trajectory, a measurement of their vertical acceleration is
performed by a Raman interferometry scheme \cite{chu:l_pulses2}.
External source masses are positioned in two different configurations and the
induced phase shift is measured as a function of masses positions. In order to
suppress common-mode noise and to reduce systematic effects, a
double-differential scheme has been adopted. The vertical acceleration is
simultaneously measured in two vertically separated positions with two atomic
samples, that are launched in rapid sequence with a juggling method.  From the
differential acceleration measurements as a function of the position of source
masses, and from the knowledge of the mass distribution, the value of
\textit{G} can be determined.

The paper is organized as follows: In Section \ref{ramansection} we introduce
Raman interferometry and the basic idea of the experiment to measure
\textit{G}. Section \ref{section:setup} describes the apparatus and the
experimental sequence.  A characterization of interferometer and gradiometer
performaces is presented in Section \ref{section:int-grad}. In Section
\ref{section:G-gradiometer} we report  the detection of external source masses.
In Section \ref{subsec:G}, we discuss the expected performance with the final
configuration and the projected accuracy in the measurement of \textit{G}.

\section{Raman interferometry basics and idea of the experiment}\label{ramansection}

In this section, we discuss the basic idea of the experiment, the scheme of
Raman interferometry and its application to measure \textit{G}. A more detailed
discussion can be found in \cite{fattori:magia} and references therein.

In a Raman interferometry-based gravimeter,  atoms in an atomic fountain are
illuminated by a sequence of three light pulses. The light pulses are realized
with two laser beams, which have frequencies $\omega_1$ and $\omega_2$ close to
transitions of a $\Lambda$-type three-level atom with two lower states
$|a\rangle$ and $|b\rangle$ and an excited state $|e\rangle$. The laser beams
propagate along the vertical $z$-axis in opposite directions and with
wavevectors $\vec k_1= k_1 \vec e_z$ and $\vec k_2= - k_2 \vec e_z$ ($k_i =
\omega_i/c,\,\rm i=1,2$). The light pulses drive two-photon Raman transitions
between $|a\rangle$ and $|b\rangle$. A $\pi$-pulse, which has a duration of
$\tau=\pi/\mathrm{\Omega}$ ($\mathrm{\Omega}$ being the two photon Rabi
frequency), switches the atomic state from $|a\rangle$ to $|b\rangle$ or
viceversa. A $\pi/2$-pulse with duration $\tau=\pi/(2\Omega)$ splits the atom
wavefunction into an equal superposition of $|a\rangle$ and $|b\rangle$.
Besides altering the real amplitudes $\alpha,\beta$ of the atomic wavefunction
$\Psi = \alpha \,{\rm e}^{i\Phi_\alpha} |a\rangle + \beta \,{\rm
e}^{i\Phi_\beta} |b\rangle$, the light field can also modify the atomic
momentum and the phase. An atom that changes its internal state receives a
momentum transfer by an amount of $\hbar\, k_{\rm eff}=\hbar\,(k_1 + k_2)$. At
the same time, the phases $\Phi_{\alpha,\beta}$ are modified according to the
local phase of the light field.

%
%
%
%

The interferometer is realized by a sequence of a $\pi/2$-pulse, a $\pi$-pulse
and $\pi/2$-pulse, each separated by the time $T$, that produces two possible
paths, ${\rm I}$ and ${\rm II}$, in space-time. The first (${\pi/2}$ pulse)
splits the atomic wave packet, the second (${\pi}$ pulse) induces the internal
and external state inversion and the third (${\pi/2}$ pulse) recombines the
matter waves after their different space--time evolution. After recombination
of the two paths, at the output of the interferometer the probability of
finding the atoms in state $|a\rangle$ shows a typical interference-like
behaviour:
\begin{equation}\label{interference}
N_a/N\propto 1-\cos (\Phi_{\rm I}-\Phi_{\rm II})\:
\end{equation}
$\Phi_{\rm I,II}$ are the phases accumulated on path I and II, respectively.
The phases $\Phi_{\rm I}$ and $\Phi_{\rm II}$ depend on the local phase of the
light field as seen by the atoms during Raman pulses. This links the vertical
atomic position to the phase evolution of the laser field. The phase evolution
depends on the effective frequency $\omega_{\rm
eff}(t)=\omega_1(t)-\omega_2(t)$ and on the phase relation between the Raman
pulses. Usually, one varies the effective frequency linearly in time to
compensate for the first order Doppler effect of the free-falling atoms and
keep the Raman resonance condition (effective frequency matching the energy
splitting of the two lower states). With $\omega_{\rm eff}(t)=\omega_{{\rm
eff},0}-\beta t$ one then obtains:
\begin{equation}\label{wecomp}
\Phi_{\rm I}-\Phi_{\rm II}=(\beta - k_{\rm eff}g) T^2-
\phi(0)+2\phi(T)-\phi(2T)\;
\end{equation}
The compensation of the Doppler effect ($\beta=k_{\rm eff}g$) and an
unperturbed evolution of the laser phase $\phi$ leads to $\Phi_{\rm
I}-\Phi_{\rm II}=0$. Actively changing the laser phase between $T$ and $2T$ by
$\delta\phi$ will result in $\phi(2T)=\phi(0)+\delta\phi=\phi(T)+\delta\phi$
and hence $\Phi_{\rm I}-\Phi_{\rm II}=-\delta\phi$. In this way, one can scan
the interference fringe to prove $(\beta - k_{\rm eff}g)=0$ (for right $\beta$)
or reveal the phase offset $(\beta - k_{\rm eff}g)T^2$ (for imperfect Doppler
compensation). In both cases, the value of $g$ is obtained combining the
measured phase offset and the value of $\beta$, which is set by a frequency
generator.

For the measurement of \textit{G}, we use a Raman interferometer to detect the
change in atoms acceleration induced by external source masses. In order to
achieve high sensitivity and accuracy, however, the experimental scheme was
developed with important specific features.

First, we launch two clouds of atoms to realize two fountains that are
displaced vertically (see Figure \ref{fig:setup}). The Raman pulses act on both
clouds simultaneously and generate two interferometers at the same time. In a
detection of the differential phase shift between the two interferometers,
spatial homogeneous accelerations cancel and common mode measurement noise is
reduced. The two-cloud setup results in an atomic gravity gradiometer. If the
trajectory of the first cloud is located above the source masses, atoms will
experience an induced acceleration in -$z$ direction. In contrast, choosing the
trajectory of the second cloud below the source masses, the induced
acceleration of this cloud will be in +$z$ direction. Taking the difference of
these two accelerations yields a signal, which is about twice the one obtained
with only one cloud.

Second, we determine the differential interferometer phase shifts with the
source masses at distinct positions (positions 1 and 2 in Figure
\ref{fig:setup}). Evaluating the difference between the measurements further
reduces systematic spurious effects if they are constant over the time scale of
the source masses repositioning.

Third, the atomic trajectories and the shape and positioning of source masses
were optimized to reduce experimental sensitivity to crucial parameters like
precision and stability of the atomic fountain.

The combination of these features will allow to reach the targeted accuracy
$\Delta G/G \approx 10^{-4}$.


\section{\label{section:setup} Experimental setup}


The experimental apparatus consists of the vacuum chamber, the laser sources
and optical set up for the production of the double fountain of Rb atoms using
a juggling procedure, the apparatus for Raman pulse interferometry, and the
system of source masses, their support and precision positioning components.
The experimental sequence, timing and data acquisition are computer controlled.
In this section, we describe the main parts of the apparatus. More details can
be found in \cite{PetelskiThesis}.


\subsection{\label{subsec:laser} Atomic fountain and juggling apparatus}

The fountain  of ${^{87}\text{Rb}}$ atoms is produced using a magneto-optical
trap and moving--molasses scheme. The laser beams are in a 1-1-1 six-beam
$\sigma^+/\sigma^-$ configuration. This keeps the central vertical axis free
for the Raman laser beams and allows to realize a stable and precise atomic
fountain.

The relevant parts of the vacuum system are shown in Figure \ref{fig:setup}. It
consists of three main parts: The lowest part is a titanium cube with cut edges
where atoms are trapped in the MOT from the vapour produced by a dispenser
(SAES Getters 5G0807), cooled and launched in the fountain.  The middle part of
the vacuum system is a thermally demagnetized 316 LN stainless steel cell. This
cell is used for pumping and to detect the atoms. The top part of the vacuum
system is the interferometer tube. It is 1~m long, has a diameter of $40$~mm
and is made of titanium. The tube is magnetically shielded with two coaxial
cylinders of $\mu$--metal (Amuneal) that are ${0.76\ \text{mm}}$ thick and ${1\
\text{m}}$ long, with internal diameters of ${74\ \text{mm}}$ and ${84\
\text{mm}}$. The attenuation of the radial and axial components of the external
field is ${76\ \text{dB}}$ and ${69\ \text{dB}}$, respectively. The saturation
field ${\text{B}_\text{S}}$ is about ${6\ \text{mT}}$.

A magnetic field gradient of 75 mT/m  in the MOT region is generated by a
water-cooled pair of coils in anti--Helm\-holtz configuration. The cooling
radiation is generated by a tapered amplifier delivering ${500\ \text{mW}}$
output power. It is injected with ${25\ \text{mW}}$ from a laser (New Focus
Vortex 6000, ${65\ \text{mW}}$ at ${125\ \text{mA}}$) frequency--stabilized on
the $5 \, ^{2}\text{S}_{1/2},$ F=2 $\rightarrow$ $5 \, ^{2}\text{P}_{3/2},$ F=3
transition of ${^{87}\text{Rb}}$. This laser also acts as the main frequency
reference for the experiment. The radiation for the MOT is detuned by
${3\mathrm{\Gamma}}$ ($\mathrm{\Gamma} = 2 \pi \cdot 6.1$ MHz) to the red of
the resonance.

Repumping light is provided by an extended cavity diode laser,
frequency--locked to the reference laser and optically injecting a slave diode
laser.

Cooling light is coupled into fibers that are connected to an integrated fiber
splitter system (Sch\"after\&Kirchhof) which distributes the  light to the $6$
MOT beams which have an intensity of 4.2 $\text{mW/cm}^2$ each with 11 mm beam
waist. Each beam is delivered to the MOT region through a collimator rigidly
fixed to the cell. Hyperfine repumping light is provided by a beam in one
direction with an intensity of 0.8 $\text{mW/cm}^2$.

Under standard operating conditions, $\sim {10^{9}}$ atoms are loaded in the
MOT with a typical loading time  ${\tau=2.5\ \text{s}}$. The MOT has a size of
$ \sim 4 $ mm.

\begin{figure}
\centering
\includegraphics[width=8.8cm]{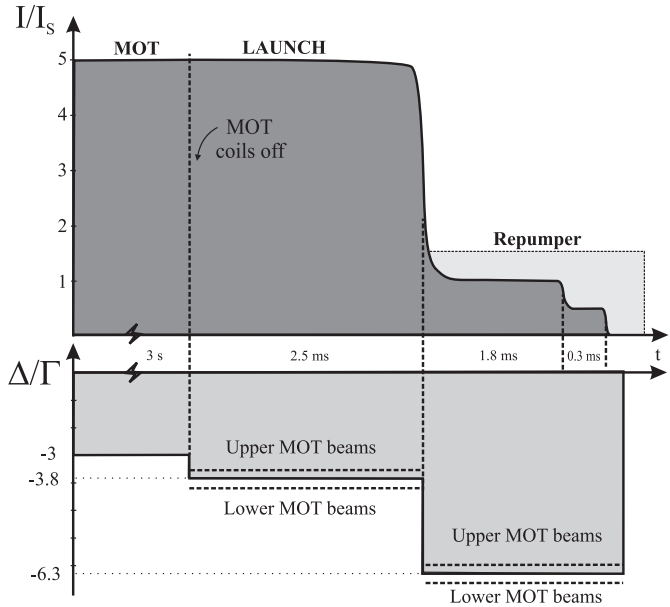}
\caption{\label{fig:launch} Sequence used in the experiment to launch cold
atoms in the fountain. Horizontal dashed lines represent the different
frequencies for the downwards- and upwards-propagating beams. The time axis is
not to scale.}
\end{figure}

The atomic cloud is launched into the fountain tube tube by using the moving
molasses method. The laser beams propagating upwards and downwards are
separately controlled by two acousto--optic modulators (AOMs). By applying a
detuning ${\mathrm{\delta}\omega}$ of opposite sign to the upper and lower
beams,  the atoms are forced to move along the vertical direction with a
velocity
\begin{equation}
v=\frac{\mathrm{\delta}\omega}{k \, \cos\xi} \label{eqn:launch_vel}
\end{equation}
where ${k}$ is the wave vector of the cooling radiation and ${\xi}$ the common
angle between each of the six  beams and the vertical direction ($\cos
\xi$=1/$\sqrt{3}$). During the launch, the detuning of the cooling beams in the
moving frame is increased and their intensity is reduced. This sequence allows
to further cool the atomic sample in the moving molasses frame. The launch
sequence for one cloud is realized in four steps (Figure \ref{fig:launch}).
After the MOT magnetic field is switched off, the atoms are launched upwards by
introducing a detuning of opposite sign to the three upwards- and
downwards-propagating laser beams. After ${2.5\ \text{ms}}$, the intensity of
the beams is lowered to the saturation intensity ${\text{I}_\text{s}} = 1.7 \,
\text{mW/cm}^2$ and the mean detuning is increased to ${-6.3 \,
\mathrm{\Gamma}}$. After ${1.8\ \text{ms}}$, the intensity of the cooling beams
is reduced by a factor of 2 for ${0.3\ \text{ms}}$. Finally, the cooling beams
are switched off, leaving only the repumper beam on to optically pump the atoms
into the ${\text{F}=2}$ state.
 The temperature
of the sample, measured by monitoring its axial expansion during free fall, is
approximately ${4\ \mu \text{K}}$. The number fluctuation from launch to launch
is about 3\%. The fountain sequence is synchronous with the 50 Hz power line.

\begin{figure}
\centering
\includegraphics[width=8.8cm]{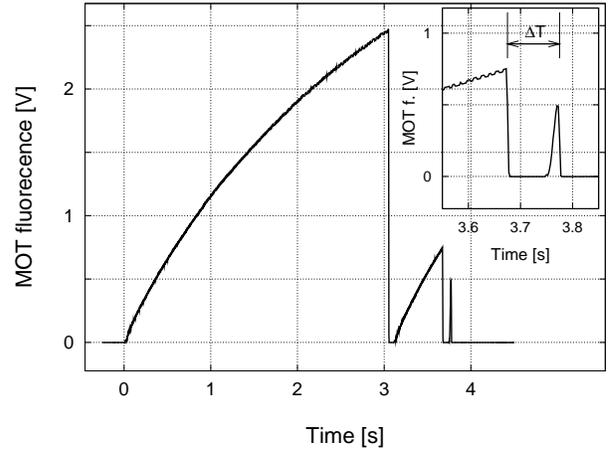}
\caption{\label{fig:juggling} MOT Fluorescence detected during a juggling
sequence. The first two clouds are slowly loaded from background gas. The third
one is obtained trapping the first cloud falling down in the MOT region. The
inset shows the launch  of the last two clouds in a time interval of 100 ms.}
\end{figure}

The gradiometer requires two clouds of cold atoms moving with the same velocity
at the same time, but vertically displaced. A vertical separation of ${35\
\text{cm}}$ for atoms launched 60 cm and 95 cm above the MOT results in a
launch delay between the two clouds of about ${100\ \text{ms}}$. The two atomic
clouds are prepared using the juggling technique \cite{gibble:juggling}. During
the ballistic flight of the first cloud of atoms a second cloud is loaded from
the backgroud vapour. Just before the first cloud falls down in the MOT region,
the second one is launched. Then, the first cloud, used as a cold and intense
source of atoms, is recaptured, cooled and launched upwards within less than
${50\ \text{ms}}$ (Figure \ref{fig:juggling}).

To optimize the juggling sequence, several factors must be taken into account.
First, the recapture efficiency of the atomic fountain decreases with the
launch height. The dominant loss process is the scattering with the thermal
background atoms. The loss process due to the thermal expansion of the cloud
during the flight is so far negligible. Second, the higher the launch of the
first cloud, the longer the loading time for the second one. Third, increasing
the time used to collect atoms for the first cloud has two opposite effects:
the number of recaptured atoms after the launch is higher, but the experimental
repetition rate is reduced, and so the instrument sensitivity.

In our experimental sequence, the first cloud is launched ${60\ \text{cm}}$
upwards, which leads to a loading time of
 650 ms for the second cloud.
In this way the number of atoms launched in each of the two clouds used in the
gradiometer is ${5 \cdot 10^8}$.

\subsection{\label{subsec:preparation} State preparation procedure}

After the launch, the atoms are selected both in velocity and by their
$\text{m}_\text{F}$ state. The selection procedure uses vertical beams so that
the state preparation can take place simultaneously on both clouds. The
selection is realized in the vacuum tube, where a uniform vertical bias field
of $25\ \mu\text{T}$ is applied. The sequence starts by blowing away the
residual population in the F=1 state, by applying a 3 ms pulse of elliptically
polarized radiation resonant with the F=1$\rightarrow$F'=0 transition. This
beam is tilted with respect to the vertical direction so that $\mathrm{\Delta
m_{F}}$=$0,\pm 1$ are possible. A narrow selection of the vertical velocity
distribution of atoms in the F=2,$\text{m}_\text{F}$=0 is transferred to the
F=1,$\text{m}_\text{F}$=0 state using a $100\ \mu\text{s}$ velocity--selective
Raman pulse.
The atoms in the F=2 state are removed from the sample by applying a circularly
polarized vertical beam for 5 ms, resonant with the F=2$\rightarrow$F'=3
transition.

After the selection sequence, the atoms end up in the F=1,$\text{m}_\text{F}$=0
state with a horizontal temperature of ${4\ \mu\text{K}}$ and a vertical
temperature of ${40\ \text{nK}}$, corresponding to velocity distribution widths
(HWHM) respectively of $3.3\ v_{\text{rec}}$ and $0.3\ v_{\text{rec}}$
($v_{\text{rec}} = 6\text{ mm/s}$ for  $\text{Rb}$ resonance transition).

\subsection{\label{subsec:interferometer} Raman interferometer apparatus}

Stimulated Raman transitions are driven by light from two extended-cavity
phase--locked diode lasers, with a relative frequency difference equal
 to the ${^{87}\text{Rb}}$ ground state hyperfine splitting frequency ($\nu_{\text{hf}}$=$6.835\
\text{GHz}$) and amplified by a single tapered amplifier. The detuning from the
D$_2$ resonance is ${-3.3\ \text{GHz}}$.

The laser locking system was described in detail in
\cite{prevedelli:phaselock}. The two main requirements of the optical
phase--locked--loop (OPLL) are robustness, necessary for continuous operation
over long periods of time, and low rms phase error
${\sqrt{\langle\varphi^2\rangle}}$, a limiting parameter for the interferometer
sensitivity. These requirements are accomplished using a detector that combines
in a mutually exclusive way a digital phase and frequency detector   and an
analog phase detector. The digital  detector allows a capture range of the
order of ${100 \,\text{MHz}}$, whereas the analog  detector ensures a low noise
spectral density necessary for accurate phase--locking.

The master laser (ML) is an extended-cavity diode laser (New Focus Vortex 6000,
${65\ \text{mW}}$ at ${130\ \text{mA}}$). The slave laser (SL) is an
anti--reflection coated diode (Sacher $\,$ SAL--0780--040, $35$ $\text{mW}$ at
$60\ \text{mA}$) mounted in an extended cavity. The ML and SL beams are
combined on a polarizing beam splitter (PBS) followed by a linear polarizer.
One output of the PBS is sent to a fast photodiode to generate the beat note.
The second PBS output injection--seeds a tapered amplifier, whose astigmatism
is corrected for with a cylindrical lens. A hot Rb cell filters resonant light
in the spectral wings of the amplifier output.

The pulse timing of the Raman beams during the interferometer sequence is
controlled by an AOM, driven by a waveform generator (Agilent 33220A). After
the AOM, the beams are coupled into a polarization--maintaining fiber, which
acts as a spatial filter. The fiber output is collimated by an aspherical lens
(f=1000 mm) to obtain a beam waist of ${10\ \text{mm}}$. Most of the optical
elements on the path of the Raman beams have a (${\lambda/20}$) quality to
avoid phase--front distorsions.

The Raman beams enter the vacuum system through the lower window of the MOT
cell and exit through the window at the top of the interferometer tube. After
passing through a quarter--wave plate, they are retroreflected by a mirror,
thus obtaining a ${lin \bot lin}$ configuration in the interferometer region.
The horizontality of the retroreflecting mirror has been adjusted within $20\
\mu \text{rad}$ using a tiltmeter (Applied Geomechanics 755--1129). Taking into
account the Doppler effect and considering the polarization of the beams, only
Raman transitions with $\mathrm{\Delta m_F}$=0 are possible for atoms with non
zero velocity. To compensate for the varying Doppler shift of the atomic
resonance during the atoms free--fall trajectory, the Raman beams frequency
difference is linearly swept using a continuous--phase waveform generator
(Agilent 33250A). The central ${\pi}$ pulse of the interferometer sequence is
sent 5 ms before the atoms reach the top of their trajectory, when their
velocity is still high enough to discriminate between upwards and downwards
propagating Raman beams. For a Raman beam intensity of ${30\ \text{mW/cm}^2}$,
the ${\pi}$ pulse lasts ${100\ \mu\text{s}}$.

The interferometric phase shifts are detected using the relative phase of the
Raman beams as a reference. To scan the interferometric fringes, a controlled
phase jump $\phi_{\text{L}}$ is applied after the ${\pi}$ pulse to the rf
signal generated by the low--phase--noise reference oscillator (Anritsu
MG3692A).

Coils wound between the vacuum tube and the magnetic shield enable a precise
control of the magnetic field during the interferometer sequence. A solenoid,
of ${1\ \text{m}}$ length, generates a vertical bias magnetic field on the
$z$--axis, chosen as quantization direction in the interferometric region. A
series of ${10}$ coils, each ${9\ \text{cm}}$ long, are used to add a
controlled phase shift via the second--order Zeeman shift on the atoms during
the interferometer sequence.

\subsection{\label{subsec:detection} Detection scheme}

After the interferometric sequence, the population of the two hyperfine
sublevels of the ground state is measured in the intermediate chamber  using
normalized fluorescence detection. The falling clouds pass through two
horizontal counterpropagating circularly--polarized beams, vertically displaced
by ${20\ \text{mm}}$. The beams have a rectangular section of ${(13 \times 6)\
\text{mm}^2}$, and their wave vector ${\vec{k}}$ is parallel to a local
magnetic field. Each beam has a power of ${1\ \text{mW}}$ and is resonant with
the F=2${\leftrightarrow }$F'=3 transition, in order to interact with atoms in
the F=2 state. The lower part of the higher beam is not retroreflected in order
to horizontally blow--away the atoms after detection. On the lower probe beam
${0.4\ \text{mW}}$ of repumping light is overlapped to optically pump atoms
from the F=1 state to the closed ${\text{F}=2 \leftrightarrow \text{F'}=3}$
transition. With this configuration, atoms in the F=2 state are detected with
the higher beam and atoms in the F=1 state with the lower one. The fluorescence
photons from the two detection regions are collected by a $ 5\ \text{cm}$
diameter lens (f=100 mm) placed at a distance of $ 130\ \text{mm}$ and
separately focused onto two large area photodiodes (Hamamatsu S7539). The
population fractions in the two states are obtained via normalization,
eliminating the shot--to--shot dependence on atom number. With a typical number
of $5\cdot 10^4$ detected atoms per cloud in each state, the signal-to-noise
ratio is $60/1$, limited by background light.

\subsection{\label{subsec:mass detection} Source masses set--up}

\begin{figure}
\centering
\includegraphics[width=7.0cm]{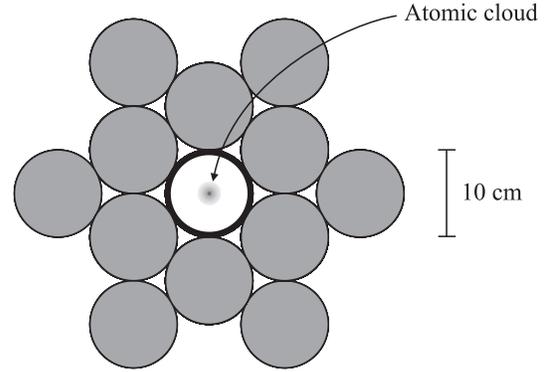}
\caption{\label{fig:hex_masses} Source masses position around the
interferometer tube (top view). }
\end{figure}

Two sets of masses are used to generate a well--known gravitational field. Each
set is made of ${12}$ identical cylinders, symmetrically arranged in a
hexagonal configuration around the vertical axis of the atomic fountain (Figure
\ref{fig:hex_masses}). The cylinders have a diameter of ${100\ \text{mm}}$ and
a height of ${150\ \text{mm}}$. The two sets of masses are held by two titanium
rings, which are connected to a precision translation stage specifically
designed for the experiment.  This allows to move vertically the two sets of
masses and to position them at a relative distance ranging between $4$ and $50\
\text{cm}$. The distance is measured with two linear optical encoders
(Heidenhain LS603). A test of the system with a laser--tracker showed a
reproducibility of $\pm 1 \mu$m.

In the final configuration for the \textit{G} measurement, well--characterized
W masses will be used (see Section \ref{subsec:G}). The results reported in
this paper were instead obtained using Pb source masses. Each cylinder had a
mass of 12.80 kg. The total mass of the 24 cylinders was 307.2 kg.

\section{\label{section:int-grad} Atom interferometer operation and characterization}


In the gravimeter configuration, the atom interferometer operating with a
single sample of atoms launched in the fountain, the main phase term is the one
induced by Earth's gravity
\begin{equation}
\phi(g)=\text{k}_{\text{eff}} \, g \, \text{T}^2 \label{eqn:phase_g}
\end{equation}
where $\hbar$k$_{\text{eff}}$ is the momentum transferred to the atoms by a
Raman pulse.

\begin{figure}
\centering
\includegraphics[width=8.8cm]{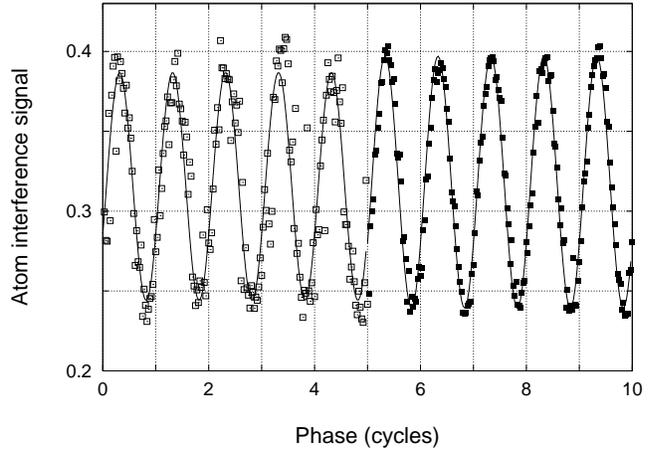}
\caption{\label{fig:fringesVibrSine} Fringes recorded with a single atom
interferometer for ${\text{T}=20\ \text{ms}}$ delay between  Raman pulses. The
active vibration isolation system of the mirror was off during the first five
cycles (empty squares) and on during the last five (filled squares). The
acquisition time for  each data set ($5$ cycles) was $900\ \text{s}$. The solid
lines are sinusoidal least--square fits over the two sets of data. The phase
uncertainty is ${33\ \text{mrad}}$ in the first case and ${10\ \text{mrad}}$ in
the second case.}
\end{figure}

The observed interference signal
is shown in Figure \ref{fig:fringesVibrSine}. The fraction of atoms in the F=2
state is plotted as a function of the phase $\phi_{\text{L}}$, which is
electronically added to the Raman lasers before the final ${\pi/2}$ pulse. The
total phase is in this case the sum of $\phi(g)$ and $\phi_{\text{L}}$. Each
data point results from a single launch of an atomic cloud ${80\ \text{cm}}$
above the center of the MOT and requires ${5.5\ \text{s}}$. The phase step is
$10^{\circ}$. The phase change due to the varying Doppler effect during the
atoms flight was cancelled by chirping the lasers' frequency.
Figure \ref{fig:fringesVibrSine} compares the signals recorded without
vibration isolation for the mirror retroreflecting the Raman beams (except for
the table air legs with active position stabilization, Newport RS 2000) and the
signal recorded when an active vibration isolation system (HWL Scientific
Instruments AVI 350--M(L)) was used to stabilize the mirror. The
signal-to-noise ratio, defined as the ratio between twice the fringe amplitude
and the rms noise, is $7$ in the first case and  $18$ in the second case. The
pulse spacing ${\text{T}}$ can be increased up to ${150\ \text{ms}}$, but the
phase noise induced by mirror vibrations drastically reduces the fringe
visibility. A good phase definition for the fringes requires a better isolation
system for the mirror and for the table where the atomic fountain is mounted.
A gravity gradient measurement is obtained by two vertically separated
acceleration measurements. The simultaneous realization of these measurements
overcomes the stringent limit set by phase noise thanks to common-mode noise
rejection. The Raman sequence interval ${\text{T}}$, as well as the gradiometer
sensitivity, can then be increased up to the limit set by experimental
constraints. If $g_{_{\text{low}}}$ and $g_{_{\text{up}}}$ are the gravity
acceleration values at the position of the lower and upper interferometers, the
differential phase shift is
\begin{equation}
\phi(\mathrm{\Delta} g)=\text{k}_{\text{eff}} \,
(g_{_{\text{low}}}-g_{_{\text{up}}}) \, \text{T}^2 \label{eqn:phase_g'}
\end{equation}
As discussed above, the gradiometer requires two clouds of cold atoms moving
with the same velocity at the same time, but vertically displaced. A vertical
separation of ${35\ \text{cm}}$ for atoms launched 60 cm and 95 cm above the
MOT requires a launch delay between the two clouds of about ${100\ \text{ms}}$.
In the present apparatus, the two atomic clouds are prepared using the juggling
technique. In the final configuration, a 2D-MOT  will be used for faster
loading of the MOT \cite{PetelskiThesis}.

\begin{figure} \centering \includegraphics[width=8.8cm]{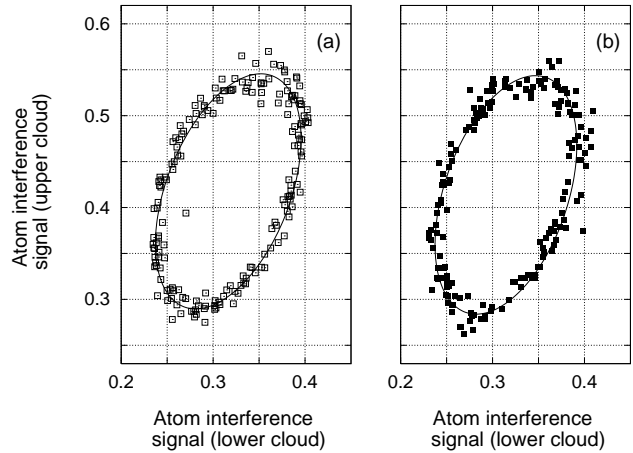}
\caption{\label{fig:fringesVibrEllipse} Signal from the gradiometer with T=20
ms. The
 fringes from the upper gravimeter are plotted versus the fringes from the
lower interferometer. The relative phase can be extracted from the elliptic fit
parameters. The mirror isolation system was on in (a), off in (b). The error on
the relative phase is  $ \sim {10\ \text{mrad}}$.}
\end{figure}
%
%
%

The Earth's gravity gradient ($\sim 3 \cdot 10^{-6}\ \text{s}^{-2}$)
corresponds to a relative phase shift that, for a vertical distance
${\mathrm{\Delta}z}$ of ${35\ \text{cm}}$ and a time interval T=100 ms between
the Raman pulses, is expected to be ${175\ \text{mrad}}$. A sensitive
measurement of the relative phase of the two interferometers is obtained by
using an ellipse fitting method \cite{kasevich:ellipsefitting} to cancel
common--mode phase--noise. The interference signal of the upper interferometer
is plotted versus that of the lower (Figure \ref{fig:fringesVibrEllipse}). The
data then describe an ellipse and the relative phase shift can be obtained from
its eccentricity and  rotation angle.  The measured signal-to-noise ratio is
about 100/1 limited by residual amplitude noise.

To adjust the operating point of the gradiometer, a controlled phase shift was
added to the lower interferometer using a magnetic field pulse, acting on the
${\text{m}_\text{F}=0}$ atoms via the second--order Zeeman effect. The
resulting phase shift can be written as
\begin{equation}
\mathrm{\Delta} \phi_{B}= 2 \pi \, \nu_{_{\text{Z,II}}}(B^2) \, \int_0^{t_{m}}
\left( B^2_{_{\text{low}}}(t) - B^2_{_{\text{up}}}(t) \right) \mathrm{dt}
\label{eqn:B_z}
\end{equation}
where $\nu_{_{\text{Z,II}}}(B^2)$=$57.515 \, \text{GHz}/\text{T}^2$ is the
second order Zeeman shift coefficient for $^{87}\text{Rb}$, ${t_m}$ is the
duration of the magnetic pulse, ${B_{_{\text{low}}}}$ and ${B_{_{\text{up}}}}$
the magnetic field values at the lower and upper accelerometer, respectively.

In this case, a knowledge of the pulsed magnetic field is needed for a precise
gradient measurement, although it is not critical in the double--differential
scheme for the $G$ measurement where only stability matters.
%
%
 The magnetic field in the interferometric region was mapped  for different current
values in the  bias coils using atoms in different ${\text{m}_\text{F}}$
sublevels as probes. A sample of cold atoms in the ${\text{F}=2}$ state was
repeatedly launched in the interferometer tube and a selected velocity class
was transferred to the F=1 state with a Raman pulse before reaching the
inversion point. The atomic sample was then transferred back to the F=2 state
with a ${\pi}$ Raman pulse and selectively detected. The delay of the second
pulse was varied, in order to interact with the free--falling atoms at
different vertical positions.
%
This measurements also provide the optimal value for the Raman beams frequency
ramp, which is $25.1354(3)\ \text{MHz/s}$ corresponding to a gravity
acceleration \textit{g} of $9.8056(1)$ $\text{m/s}^2$.
\begin{figure} \centering \includegraphics[width=8.8cm]{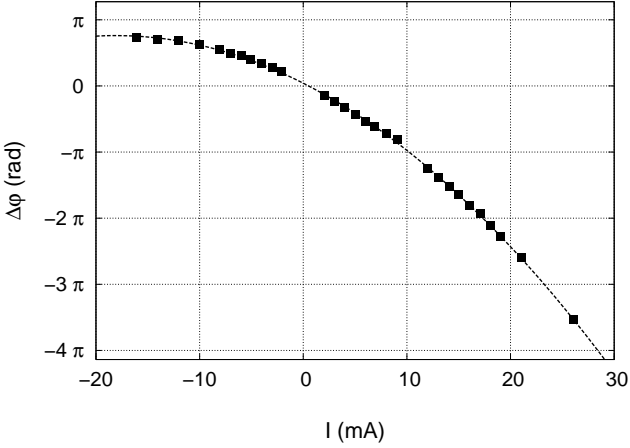} \caption{\label{fig:B_pulse}
Phase shift induced  by applying a magnetic field pulse with different
amplitudes to the lower interferometer for ${\text{T}=100\ \text{ms}}$ delay
between Raman pulses.}
\end{figure}
In Figure \ref{fig:B_pulse} a direct measurement of the magnetically--induced
phase shift is shown. Each point results from an elliptic fit on a set of
gradiometric measurements for a certain current value in a bias coil. The
parabolic fit does not cross the axis origin but is shifted by 125 mrad. This
vertical offset shoud be induced by the Earth's gradient that is constantly
present during the whole set of measurements. This result differs from the
value of 175 mrad  produced by the Earth's gravity gradient. The difference can
be attributed to the Coriolis effect.

For an atomic velocity component ${\bf{v}}_{_{EW}}$ along East-West direction,
 the phase
induced by the Coriolis effect is
\begin{equation}
\mathrm{\Delta} \phi_{C}= 2\ \bf{\Omega} \cdot (\bf{v}_{_{\text{EW}}} \times
\text{\textbf{k}}_{\text{eff}})\text{T}^{2} \label{eqn:coriolis}
\end{equation}
where $\bf{\Omega}$ is the Earth's rotation frequency ($\bf{\Omega}$ = 7.29
$\times 10^{-5}$ $\text{ rad/s}$). At the lab location (latitude
$\theta_{lat}$=43$^{\circ}$47'), a {\bf{v}}$_{\text{EW}}$ component of 1 cm/s
introduces a $\mathrm{\Delta} \phi_{C}$ corresponding to an acceleration
$\mathrm{\Delta} a=2\ {\bf{\Omega}} \text{\bf{v}}_{_{\text{EW}}}
\cos{\theta_{lat}} =10^{-7}\ g$.

In the same conditions, the resulting signal for the gradiometer configuration
is proportional to the horizontal velocity difference between the two atomic
samples:
\begin{equation}
\mathrm{\Delta} \phi_{C}= 2 \bf{\Omega} \cdot (\bf{\Delta}
\text{\textbf{v}}_{_{\text{EW}}} \times
\text{\textbf{k}}_{\text{eff}})\text{T}^{2} \label{eqn:coriolis}
\end{equation}
 For launch heights of 60 and 95 cm above the MOT, the resulting phase difference
 corresponds to
$2\cdot10^{-8} g$. This effect can be minimized by a velocity selection of the
atoms launched in the interferometer \cite{peters:metrologia}.

For the determination of \textit{G}, the gradiometer signal will be detected as
a function of the source masses positions so that rotational contributions
cancel and only fluctuations of the launch direction and height within the
measurement time can affect the signal.

\section{\label{section:G-gradiometer}  Detection of source masses}

In order to test the apparatus and to show the possibility of detecting the
effect of external source masses, the interferometer was operated in
gradiometer configuration and a set of Pb masses was used.

Pb cylinders had the shape and arrangement as described in Section
\ref{subsec:mass detection}.
\begin{figure}
\centering
\includegraphics[width=8.8cm]{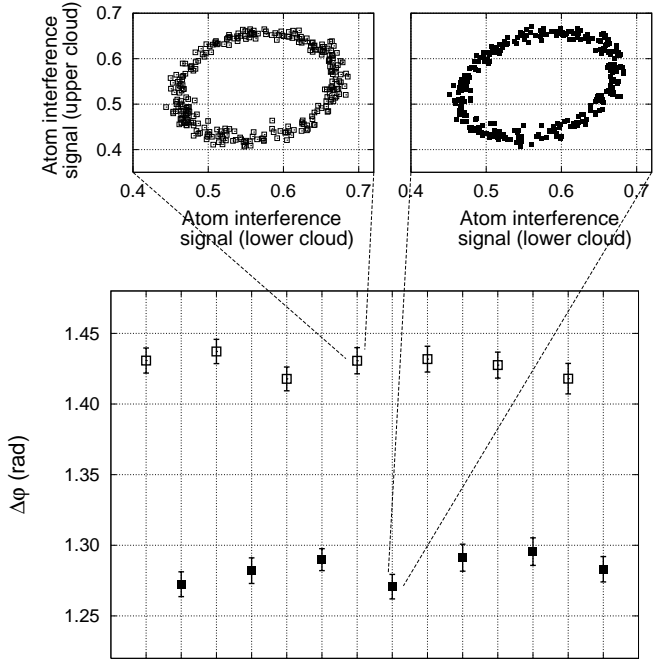}
\caption{\label{fig:masses}  Gravitational phase shift detected for the two
positions of the source masses. The experiment was performed with Pb masses and
data were analyzed using the ellipse fitting method. Each data point results
from a fit over 288 sequences with a phase step of ${5^{\circ}}$ for the local
oscillator.  The acquisition time for each point is $20$ minutes. The
translation of the source masses between two consecutive measurements required
$\sim {100\ \text{s}}$. The full data set for two measurements in different
configurations are also shown in the upper part of the figure.}
\end{figure}
Masses were alternately set in the positions corresponding to the configuration
(1) and the configuration (2) shown in Figure \ref{fig:setup}. In the first
case, the acceleration induced on the atoms in the upper cloud is in the $-z$
direction and the acceleration for the lower cloud is in the $+z$ direction.
The sign of the induced acceleration is changed, moving the masses to the
configuration (2) with respect to the atomic clouds.

In Figure \ref{fig:masses}, the differential phase shifts measured for the two
sets of Pb cylinders in the two configurations are reported. Considering the
differences between two consecutive measurements, the resulting phase shift
from the whole data set is ${144(5)\ \text{mrad}}$, which corresponds to a
sensitivity of ${3 \cdot 10^{-9}g}$ and a relative uncertainty of $4 \cdot
10^{-2}$ in the measurement of $G$. The total data acquisition time was about
${5\ \text{h}}$.

By evaluating the difference of  consecutive measurements, a reduction of
systematic effects, due for instance to spatially inhomogeneous spurious
accelerations, Earth's gravity gradient, inhomogeneous electric and magnetic
fields and inertial forces, is achieved.

\section{\label{subsec:G} Towards the measurement of \emph{G}}

To reach the planned accuracy of $\mathrm{\Delta}G / G$=$10^{-4}$ in the
measurement of \textit{G}, an optimization and characterization of the
interferometer and  source mass apparatus are required.


The source masses material for the \textit{G} measurement is a sintered alloy
made of tungsten ($95.3\%$), nickel ($3.2\%$) and copper ($1.5\%$)
 (Plansee INERMET IT180). This material is non--magnetic and has a typical
density of $(18000\pm200)\ \text{kg/m}^3$.  The 24 cylinders for the experiment
come out of the same furnace run.  Then the cylinders undergo a hot isostatic
pressing (HIP) treatment at T=${1200^{\circ} \text{C}}$, P=${1000\ \text{bar}}$
which improves the homogeneity of the material.

Characterization tests on this material were performed before and after the HIP
treatment. Samples were observed by a microscope to detect internal holes;
holes, with a typical diameter of $\sim 150\ \mu\text{m}$, were reduced to a
negligible size by the HIP treatment. The material mean density was measured
through a double weighting (in air and in water) at constant temperature. An
ultrasonic test was performed to study its homogeneity that also showed a
significative improvement after the HIP procedure. We measured the dimensions
and deviations from the cylindric shape and verified that they can be machined
with a precision of 1-2 ${\mu\text{m}}$. A sample cylinder was cut into blocks
of ${(25 \times 25 \times 44)\ \text{mm}^3}$ to characterize the internal
density distribution. The results of all tests show that masses should not
affect the final precision planned for the $G$ measurement.
\begin{figure}
\centering
\includegraphics[width=8.8cm]{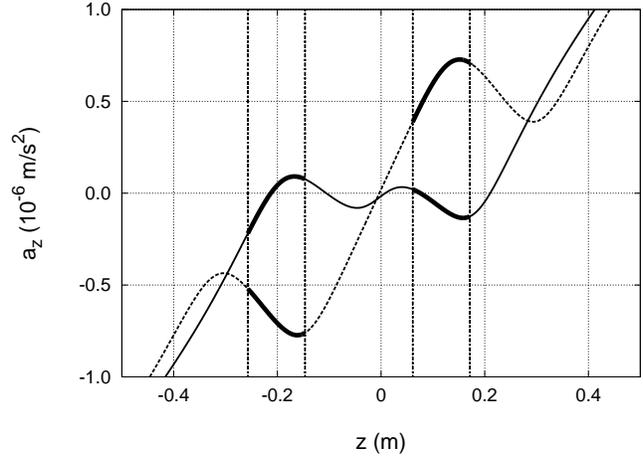}
\caption{\label{fig:acc} Vertical acceleration computed for the final
experiment  with tungsten source masses in two positions. The Earth's gravity
gradient was considered. The solid line corresponds to configuration 1 and the
dashed line to configuration 2.  The spatial regions corresponding to the
atomic trajectories for the two interferometers are highlighted.}
\end{figure}

The parameters of the atom interferometer must be optimized considering the
shape of the gravitational field produced by the source masses. For the
configuration 1 in Figure \ref{fig:acc}, selected atomic trajectories will
maximize the phase difference between the two interferometers. Keeping the same
atomic trajectories, the source masses will be moved to configuration 2 in
which the phase difference term has minimum sensitivity to the atomic position.
In this way, the interferometers will be realized where the acceleration is
stationary for both configurations. $\mathrm{\Delta}G/G$=$10^{-4}$ can be
reached assuming for the atoms a uncertainty of 1 mm for the position and 5
mm/s for the velocity. Higher levels of precision and accuracy are imposed on
the knowledge of the distance between the source masses (10  $\mu$m) and of the
relative positioning (100 $\mu$m). This will be achieved by a combination of
stable mechanical positioning and translation mount and direct optical
measurement of the position of the cylinders during the experiment.

\section{\label{section:conclusions} Conclusions}
The apparatus developed to measure $G$ using a new scheme based on atom
interferometry was presented. The atom interferometer is used to measure the
gravitational field produced by source masses. A differential scheme, realized
by an atom juggling procedure, was implemented working as a
gravity-gradiometer, thus reducing common-mode noise. The result of a
preliminary experiment was reported, where the effect of Pb source masses was
clearly detected. Further improvements in the Rb atomic fountain, a reduction
of noise sources and the use of well characterized W source masses will allow
to reach the projected precision in the measurement of $G$.

\begin{acknowledgement} This work was supported by INFN (MAGIA experiment), MIUR, ESA, and EU
(under contract RII3-CT-2003-506350). G.M.T. acknowledges seminal discussions
with M. A. Kasevich and J. Faller. We are grateful to B. Dulach of INFN-LNF for
the design of the support for the source masses and to A. Peuto of INRIM for
density tests on W masses. \end{acknowledgement}

\bibliographystyle{epj}

\end{document}